\documentclass[11pt]{article}
\usepackage{amsmath,amsfonts,amssymb,amsthm}
\usepackage{color}

\usepackage[dvipdfmx]{graphicx,xcolor}
\usepackage{hyperref}
\usepackage[margin=2cm]{geometry}
\usepackage{xspace}
\usepackage{enumitem}
\usepackage{comment}
\usepackage[utf8]{inputenc}
\usepackage[T1]{fontenc}
\usepackage{CJKutf8}
\usepackage{algorithm, algpseudocode}
\usepackage{xcolor}

\newcommand{\encode}{\mathrm{encode}}

\newcommand{\hshr}{h_{\mathrm{shr}}}
\newcommand{\hact}{h_{\mathrm{act}}}

\newcommand{\WW}{\textsc{Werewolf}\xspace}

\newcommand{\GG}{\textsc{Game of the Generals}\xspace}

\newcommand{\card}[1]{%
  \fbox{%
    \makebox[0.5em][c]{% 横幅固定
      \rule{0pt}{0.4ex}% 縦高さ固定（これが基準）
      \resizebox{!}{1.5ex}{\ensuremath{#1}}%
    }%
  }%
}

\newcommand{\questioncard}{\card{?}}
\newcommand{\heart}{\card{{\heartsuit}}}
\newcommand{\club}{\card{\clubsuit}}

\newcommand{\dia}{\card{\diamondsuit}}

\newcommand{\qcard}{
  \raisebox{-0.2ex}{\fontsize{12pt}{12pt}\selectfont \questioncard}}
\author{Naoki Kitamura, Hironori Kiya, Hirotaka Ono}
\date{}

\title{%Trust-Free Imperfect-Information Games via Physical Cryptography\\
A Moderatorless Protocol for \WW
}

\begin{document}

\maketitle
\begin{abstract}
Social deduction games, or hidden-role games, are multiplayer games in which players are assigned private roles and act under asymmetric information about other players' roles and actions. 
In the canonical example \WW, werewolves conceal their roles and mislead the other players, while the seer can obtain role information about a chosen player. 
Thus, a central functionality of such games is controlling which players can access which information. 
In typical play, this control is implemented by a trusted human moderator, who assigns roles, mediates secret actions, and reveals outcomes. 
This reliance raises the barrier to participation and introduces a trusted third party as a single point of failure.
In this work, we show that \WW can be played without a moderator or any digital device, using only ordinary playing cards. 
Our construction maintains a shared pool of cards that is observable to all players and manipulated according to a common public procedure, while its interpretation depends on each player's private role. 
This induces role-dependent views from a single public sequence of card operations. 
Consequently, even without private messages, werewolves can identify one another and coordinate, and the seer can test whether a chosen player is a werewolf in each round.
The proposed implementation is built from card-based physical cryptographic primitives, such as face-down commitments and verifiable shuffles, and higher-level subprotocols for intra-role information sharing, secret action designation, and attribute testing. 
These subprotocols implement the moderator's core functions while keeping all card operations public and auditable under standard assumptions on physical card operations. 
Several rule-violating inputs that arise naturally in the protocol, such as
selecting multiple targets, can be detected by simple consistency checks
with only a constant-factor increase in the number of cards.
We show that the resulting complete moderatorless implementation of \WW scales to an arbitrary number $n$ of players using $O(n^3)$ cards. 
We further discuss how this approach extends beyond Werewolf, suggesting
card-based ways to reduce reliance on trusted intermediaries in settings
such as \GG and second-price auctions.
\end{abstract}
%厳密には嘘．
\section{Introduction}
\subsection{Background and Motivation}
Many interactive settings rely on a trusted coordinator to manage hidden information, enforce rules, and reveal outcomes, but such reliance introduces both practical barriers and a single point of failure.
Social deduction games, or hidden-role games, are multiplayer games in which players are secretly assigned roles and act under asymmetric information. 
Players infer the identities, intentions, or actions of others from public discussion and information available only to particular players. 
A representative example is \WW, also known as Mafia. 
In \WW, players are divided into a villager faction and a werewolf faction, possibly with additional special roles such as a seer. 
As day and night phases alternate, players discuss, vote, and perform role-specific actions in order to eliminate the opposing faction. 
In typical play, a human moderator is indispensable. 
The moderator privately assigns roles, individually delivers hidden information, coordinates covert actions, and publicly announces outcomes~\cite{elias2012characteristics}.

At first glance, removing the moderator from such a game appears fundamentally difficult. 
Without a trusted third party, it is unclear how secret roles can be assigned, how information needed for each player's reasoning can be delivered only to the intended player, and how hidden actions can be coordinated without leaking sensitive information. 
Existing digital solutions, such as smartphone applications and online platforms, address these problems only partially, since they reintroduce centralized control and dependence on technical infrastructure.

In this work, we show that this dependence on a trusted moderator can be eliminated by a card-based protocol using only ordinary playing cards. 
We introduce a physical cryptographic framework in which players jointly manipulate a shared pool of cards through verifiable shuffles, commitments implemented by face-down cards, and publicly observable procedures. 
Somewhat counterintuitively, a single visible sequence of card operations can provide different information to different roles. 
By combining the shared physical state with their private role cards, players obtain role-dependent views that reproduce the information asymmetries required by the game.

This framework enables the execution of all essential components of \WW without requiring private communication or electronic devices. 
Werewolves can identify one another and coordinate their actions, the seer can test a chosen player in each round, and villagers can verify the consistency of public outcomes. 
All operations are performed sequentially under the observation of all participants, thereby making card-level deviations detectable. 
%Moreover, under standard assumptions on physical card operations, deviations that could yield a strategic advantage can be suppressed by a simple soundness amplification technique with only a constant-factor increase in the number of cards. 
Moreover, although our confidentiality analysis is in the semi-honest model, several rule-violating inputs that arise naturally in the game, such as selecting multiple targets when only one target is allowed, can be detected by adding simple consistency checks with only a constant-factor increase in
the number of cards. 
The framework generalizes to an arbitrary number $n$ of players. 
More precisely, if $h_{\max}$ denotes the maximum size of the relevant information and processing bundles, 
a direct implementation uses $O(h_{\max} n^2)$ cards per player, and hence $O(h_{\max} n^3)$ cards in total. 
In typical plays of \WW, $h_{\max}$ is constant, and this bound becomes $O(n^3)$ in total.  
Moreover, by reusing cards via a copy primitive for commitments, the number of distinct physical cards can be reduced to \(O(h_{\max} n^2)\), 
or $O(n^2)$ in typical cases. 
%For clarity, we describe a direct implementation in which each player uses $O(n^2)$ cards, i.e., $O(n^3)$ cards in total. 
%The total number of cards can be reduced to $O(n^2)$ by reusing cards via a copy primitive for commitments.
%For constant-size information bundles, 
%The number of distinct physical cards can be reduced to \(O(n^2)\) by reusing cards via a copy primitive for commitments.
This paper gives a complete moderatorless protocol for Werewolf and discusses
how the same viewpoint may apply to other imperfect-information settings
that require centralized information control.
As examples, we consider \GG, a board game with hidden pieces and referee-mediated
combat resolution, and second-price auctions,
also known as Vickrey auctions. 
Thus, the significance of our approach is not limited to recreational games: it suggests that even interactions involving complex hidden information can be implemented as low-tech secure multi-party procedures with reduced reliance on trusted intermediaries.
In this sense, our work also offers a perspective on decentralized coordination, participatory systems, and human-centered cryptography.

\subsection{Related work and our contributions}
Card-based cryptography originates from the five-card trick of den Boer~\cite{denBoer1989}, which showed that two players can compute the AND of their private bits using only a small deck of physical cards. 
Crépeau and Kilian~\cite{Crepeau1994} subsequently developed a broader theoretical foundation for card-based protocols, including protocols for generating hidden random permutations, and demonstrated the general potential of physical cards as a medium for secure computation. 
Since then, card-based cryptography has largely focused on secure computation of basic functions and related cryptographic tasks using physical cards. 
More recently, however, a line of work has begun to apply card-based cryptographic techniques directly to games. 
Shinagawa et al.~\cite{shinagawa2025play} introduced card-based protocols for playing Old Maid with virtual players, where the main task is to remove pairs from a virtual player's hand without revealing unnecessary information about that hand. 
Ruangwises and Shinagawa~\cite{ruangwises2025simulating} extended this direction to UNO by designing a protocol that simulates virtual players: in each turn, the protocol selects a valid card uniformly at random from a virtual player's hand, or reports that no valid card exists, without revealing the rest of the hand. 
Ikeda and Shinagawa~\cite{ikeda2025play} considered another closely related direction, showing how to play Mastermind without a game master by replacing the game master's feedback and code-generation functions with card-based protocols.

These works are close in spirit to ours in that they use physical cryptographic protocols not merely to compute isolated Boolean functions, but to implement game mechanisms that would otherwise require hidden information or an external entity. 
A particularly relevant prior work is the secure grouping protocol of Hashimoto et al.~\cite{hashimoto2018secure}. 
Their protocol addresses secure group assignment and has been noted as applicable to assigning character roles in \WW. 
As observed in the related-work discussion of Shinagawa et al.~\cite{shinagawa2025play}, however, such an application would cover only the role-assignment part, and would not make all parts of \WW playable without a moderator.

Our paper takes a different and broader step. 
While secure grouping captures an important role-assignment component, a full moderatorless implementation of \WW requires more than allowing players with the same role to recognize one another. 
The moderator continuously controls role-dependent information during play: werewolves must recognize and coordinate with one another, the seer must privately learn information about a chosen player, hidden night actions must be resolved, and public outcomes must be announced in a verifiable way. 
We address these requirements by introducing higher-level subprotocols for intra-role information sharing, secret action designation, and attribute testing, and by combining them into a complete moderatorless implementation of \WW.

\medskip

The rest of this paper is organized as follows. 
Section~2 introduces the model of card-based protocols and the basic card operations used in our construction. 
Section~3 presents the moderatorless protocol for Werewolf, together with
extensions to additional roles and an efficiency analysis. 
Section~4 discusses Game of the Generals and second-price auctions as examples
illustrating how the same card-based viewpoint can be used to reduce
reliance on a referee or trusted information manager beyond \WW.
Section~5 concludes the paper and outlines possible directions for future work.

\section{Preliminaries}
In this section, we introduce the model of card-based secure computation used in this work, and define the basic card-based primitives that serve as building blocks for our later constructions.

\subsection{Model}
We model card-based secure computation using physical decks of cards. 
We use two types of cards: suit cards, whose front side shows one of the four suits
(clubs, hearts, diamonds, spades), and numeric cards, whose front side shows an
integer label. 
When the number of players is $n$, we assume that numeric cards with labels
$0,1,\ldots,n$ are available.
For both types of cards, all backs are identical and reveal no information about
the front side.
%Strictly speaking, all our protocols can be implemented using only numeric cards labeled $0$ and $1$, but we introduce suit cards for clarity of exposition. Thus, every card used in this paper is either a suit card $\clubsuit,\heartsuit,\diamondsuit,\spadesuit$ or a numeric card labeled with an integer between $0$ and $n$.
The suit cards are used only for readability in binary commitments and may be
replaced by two distinguished numeric labels, such as $0$ and $1$.

\begin{comment}
%直訳調

In this paper, we realize secure computation using physical decks of cards.
We use multiple copies of two kinds of cards: cards whose front side
shows one of the four suits (clubs, hearts, diamonds, and spades) and
cards whose front side shows an integer from $0$ to $n$. The backs of
all cards are indistinguishable. Strictly speaking, it would be possible
to implement all of our protocols using only the cards labeled $0$ and $1$, but we use the suit cards as well for ease of explanation.
That is, each card has on its front side either a suit symbol
$\clubsuit,\heartsuit,\diamondsuit,\spadesuit$ or a number between $0$
and $n$, and all backs have the same design so that they cannot be
distinguished from one another.

\end{comment}

\subsubsection{Commitments}
We represent a single bit $a \in \{0,1\}$ by an ordered pair of two cards
chosen from $\clubsuit$ and $\heartsuit$.
We use the order $\clubsuit\,\heartsuit$ to encode $0$ and the order
$\heartsuit\,\clubsuit$ to encode $1$, and write
\[
\encode(0) = (\clubsuit,\heartsuit), \qquad
\encode(1) = (\heartsuit,\clubsuit).
\]
A \emph{commitment} to a bit $a$ is a face-down pair of cards placed
according to $\encode(a)$.
Unless the protocol explicitly reveals the pair, no observer can learn the value of $a$.
When verification is required, for example at the end of a protocol or at the end of a game of \WW, the pair can be turned face-up to reveal and check the committed bit.
%As long as the cards remain face down, no observer can learn the value of $a$, but at the end of a protocol (for example, at the end of a game of Werewolf) the pair can be turned face-up to reveal and verify the committed bit.

\subsubsection{Basic operations}\label{subsubsec:operaration}
We describe card movements by permutations written in one-line notation.
A permutation $\sigma$ on $n$ positions is written as
\[
(\sigma(1), \sigma(2), \dots, \sigma(n)),
\]
and represents the operation that moves the card currently in position
$i$ to position $\sigma(i)$.
For example, $(1,3,4,2)$ denotes the permutation with
$\sigma(1) = 1$, $\sigma(2) = 3$, $\sigma(3) = 4$, and $\sigma(4) = 2$.
We write $\mathsf{id}$ for the identity permutation, which leaves all
positions unchanged.
%For brevity, we collectively refer to the randomized permutation operations above as shuffles.

We now specify the elementary operations that our protocols are allowed to use.
All of these operations are performed under the joint observation of all players.
\begin{itemize}
\item \textbf{Swap}:
Swap the positions of two specified cards while keeping them face-down.
Swapping positions $i$ and $j$ with $i < j$ corresponds to the
permutation
\[
(1,\dots,i-1, j, i+1,\dots,j-1, i, j+1,\dots,n).
\]
\item \textbf{Random Cut}:
Cut a pile of $n$ cards into two parts at a uniformly random position
$k \in \{0,1,\dots,n-1\}$ and then swap the top and bottom parts.
This applies the permutation
\[
(k+1, k+2, \dots, n, 1, 2, \dots, k),
\]
which is a global rotation that preserves the relative order of the
cards.
Applying the random-cut operation repeatedly yields the same
distribution as a single random cut with a uniformly random shift.
\item \textbf{Random Bisection Cut (RBC)}:
Given a pile of $2m$ cards, cut it exactly in half and either leave the two halves in place or swap them, each with probability $1/2$. Equivalently, with probability $1/2$ we apply the identity permutation $\mathsf{id} = (1,2,\dots,2m)$, and with probability $1/2$ we apply
\[
(m+1, m+2, \dots, 2m, 1, 2, \dots, m).
\]
  \item \textbf{Pile-Scramble Shuffle:}
  Given $d$ piles of equal size, which we denote by $P_1,\ldots,P_d$, sample
  a uniformly random permutation $\sigma \in S_d$ that remains unknown to
  all players, and reorder the piles so that the $i$-th pile becomes $P_{\sigma(i)}$.
In most applications in this paper, after applying a pile-scramble
shuffle to a collection of piles, we place the shuffled piles back into
the same set of positions from which they were gathered, using their
new random order.  Thus, when the piles were gathered from rows,
columns, or grid positions, the shuffled piles are arranged again as
rows, columns, or grid positions, respectively.

  \item \textbf{Pile-Shifting Shuffle:}
  Given $d$ piles of equal size, which we denote by $P_1,\ldots,P_d$, sample
  a uniformly random offset $r \in \{0,1,\ldots,d-1\}$ that remains unknown
  to all players, and reorder the piles so that the $i$-th pile becomes
  $P_{\sigma(i)}$, where
  \[
    \sigma(i) = ((i + r - 1) \bmod d) + 1.
  \]
  In other words, the piles are cyclically shifted by a hidden random offset.
  A physical implementation is to bind each pile with an identical rubber
band and apply a random cut to the whole sequence of piles, as in the
Random Cut operation.
As with the pile-scramble shuffle, in most applications in this paper,
after applying a pile-shifting shuffle to a collection of piles, we
place the shifted piles back into the same set of positions from which
they were gathered, using their new cyclic order.

%\item \textbf{Pile-Scramble Shuffle}:
%Given $d$ piles of equal size $(\text{pile}_1,\dots,\text{pile}_d)$, reorder the piles according to a uniformly random permutation $\sigma \in S_d$ that remains unknown to all players. After the shuffle, the order becomes $(\text{pile}_{\sigma(1)},\dots,\text{pile}_{\sigma(d)})$. A physical implementation is to place each pile in an envelope and have the players jointly mix the envelopes on the table.
%\item \textbf{Pile-Shifting Shuffle}:
%Given $d$ piles of equal size $(\text{pile}_1,\dots,\text{pile}_d)$, cyclically shift their order by a uniformly random offset $r \in \{0,1,\dots,d-1\}$ unknown to all players. After the shuffle, the $i$-th pile becomes $\text{pile}_{(i+r \bmod d)+1}$.
\item \textbf{Reveal}: 
Turn one or two specified cards face-up so that all players can inspect
their faces.
\item \textbf{Discard}:
Remove specified cards from the game, moving them to a location that no
player can access thereafter.
Cards may be discarded either after being revealed or while still
face-down.
\item \textbf{Confirm}:
Allow only a designated player to privately inspect the faces of one or
two specified cards (for example, the outputs of a shuffle that have not
been revealed).
All players see which cards are inspected and by whom, but only the
designated player learns their contents.
\end{itemize}

\subsubsection{Security model}
In our setting, physical operations on cards---such as placing,
revealing, and shuffling---are executed sequentially under the joint
observation of all players.
As a result, any deviation at the level of these operations can be
detected while the protocol is running.
We therefore treat such deviations as disallowed and assume that each
player follows the prescribed protocol steps exactly as specified.
However, we do allow each player to use all information she can observe
during the execution in an attempt to infer other players' secret
inputs.
The security model used to analyze confidentiality under this
assumption is the \emph{semi-honest model} (also known as the
honest-but-curious model)~\cite{Goldreich2004}.

This semi-honest model applies to card operations and not to strategic
decisions in game play.
In particular, we do not rule out the possibility that a player
supplies an input that is forbidden by the game rules.
For example, a player might select several candidates at once in a
situation where she is supposed to designate a single player.
Our protocols are designed so that such inputs appear as violations of
consistency conditions during execution.
A detailed analysis of how to detect these invalid inputs and how the
game should respond to them lies outside the main scope of the security
model considered in this paper.

\subsection{Primitives}\label{sec:primitives}
In this subsection, we collect the card-based building blocks for
Boolean computation that we use throughout the paper.
Conventionally, these constructions are described as ``card-based
protocols'' for basic operations such as AND and XOR, but in our
framework we treat them as reusable primitives that can be composed
into more complex protocols, such as our moderatorless Werewolf
protocol in Section~\ref{sec:werewolf}.

\subsubsection{AND}\label{sec:and}

We use an \emph{AND primitive} that takes as input commitments to two
bits $a,b \in \{0,1\}$ and outputs a commitment to $a \land b$.
This primitive is implemented by the six-card AND protocol of Mizuki
and Sone~\cite{MizukiSone2009}, instantiated in our shuffle model using
only the allowed operations (random bisection cut, permutations,
reveal, and discard).
We index the positions in a row of cards from left to right by
$1,2,3,4,5,6$.
We describe card movements by permutations written in one-line notation as in Section~\ref{subsubsec:operaration}.
In particular, a permutation $\sigma$ on $n$ positions is written as
$(\sigma(1),\ldots,\sigma(n))$ and represents the operation that moves
the card currently in position $i$ to position $\sigma(i)$.

\begin{enumerate}
\item \textbf{Initial arrangement}\\
We prepare commitments $[a]$ and $[b]$ to the input bits $a$ and $b$,
and a fixed commitment $[0]$.
From left to right, we place the two-card encodings of $a$, $b$, and
$0$ in this order:
\[
\begin{array}{rclrclrcl}
\qcard & \qcard & & \qcard & \qcard & & \club & \heart & \\
\multicolumn{2}{c}{\mathrm{encode}(a)} & &
\multicolumn{2}{c}{\mathrm{encode}(b)} & &
\multicolumn{2}{c}{\mathrm{encode}(0)} &
\end{array}
\]
Thus the six cards are, from left to right,
\[
(c_1, c_2, c_3, c_4, c_5, c_6)
= (\text{first card of $[a]$}, \text{second card of $[a]$},
   \text{first card of $[b]$}, \text{second card of $[b]$},
   \clubsuit, \heartsuit),
\]
where we use $(\clubsuit, \heartsuit) = \mathrm{encode}(0)$ and
$(\heartsuit, \clubsuit) = \mathrm{encode}(1)$.

\item \textbf{Apply the first permutation $\sigma_1$.}
We apply the permutation
\[
  \sigma_1 = (1,4,2,3,5,6)
  \quad\Longleftrightarrow\quad
  \begin{pmatrix}
    1 & 2 & 3 & 4 & 5 & 6 \\
    1 & 4 & 2 & 3 & 5 & 6
  \end{pmatrix}
\]
to the six cards.
The card in position \(1\) remains in position \(1\), the card in
position \(2\) moves to position \(4\), the card in position \(3\) moves to
position \(2\), the card in position \(4\) moves to position \(3\), and the
cards in positions \(5\) and \(6\) remain in positions \(5\) and \(6\),
respectively.
Thus the sequence of cards changes from
\[
(c_1,c_2,c_3,c_4,c_5,c_6)
  \longrightarrow
(c_1,c_3,c_4,c_2,c_5,c_6).
\]

\item \textbf{Apply a random bisection cut} \\
We split the six cards into a left block consisting of positions
$1,2,$ and $3$ and a right block consisting of positions
$4,5,$ and $6$, and apply one random bisection cut to these
two blocks.
Equivalently, we sample a uniformly random bit $r \in \{0,1\}$ and
do the following:
\begin{itemize}
  \item if $r = 0$, we leave the order of the cards unchanged
        as $(1,2,3,4,5,6)$;
  \item if $r = 1$, we swap the left and right blocks, obtaining
        $(4,5,6,1,2,3)$.
\end{itemize}
We interpret the outcome of this RBC as implicitly generating the
uniform random bit~$r$.

\item \textbf{Second permutation $\sigma_2$.}
We then apply the permutation
\[
 \sigma_2 = (1,3,4,2,5,6)
  \quad\Longleftrightarrow\quad
  \begin{pmatrix}
    1 & 2 & 3 & 4 & 5 & 6 \\
    1 & 3 & 4 & 2 & 5 & 6
  \end{pmatrix}
\]
to the six cards resulting from the RBC.
In other words, the card in position \(1\) remains in position \(1\), the card in position \(2\) moves to position \(3\), the card in position \(3\) moves to position \(4\), the card in position \(4\) moves to position \(2\), and the cards in positions \(5\) and \(6\) remain in positions \(5\) and \(6\), respectively.

After these operations (the first permutation, the RBC, and the second permutation), the final sequence of cards $(d_1,\ldots,d_6)$ is determined by the values of $a$, $b$, and $r$.

\item \textbf{Revealing the first two cards and selecting the output}\\
We turn the cards at positions $1$ and $2$ face-up and decide the output
positions according to their pattern (encoding $0$ or $1$):
\begin{itemize}
\item If positions $1$ and $2$ show $\clubsuit\,\heartsuit$ (encoding
$0$), we take the cards at positions $5$ and $6$ as the commitment to
$[a \land b]$ and discard the cards at positions $3$ and $4$.
\item If positions $1$ and $2$ show $\heartsuit\,\clubsuit$ (encoding
$1$), we take the cards at positions $3$ and $4$ as the commitment to
$[a \land b]$ and discard the cards at positions $5$ and $6$.
\end{itemize}
Since $r$ is uniformly random, the pattern revealed at positions $1$ and
$2$ leaks no additional information about the input bits $a$ and $b$.

\paragraph{Correctness.}
We now verify that the protocol always outputs a correct commitment to
$a \land b$.
Table~\ref{tab:and-correctness} lists, for each choice of
$a,b \in \{0,1\}$ and $r \in \{0,1\}$, the pattern revealed at positions
$1$ and $2$, the positions chosen as output, the corresponding pair of
cards, and the decoded output bit.

\begin{table}[h]
\centering
\begin{tabular}{c c c c c c c}
$a$ & $b$ & $r$
  & cards at pos.\ 1,2
  & output pos.
  & output pair
  & output bit \\
\hline
0 & 0 & 0 & $\clubsuit\,\heartsuit$ & 5,6 & $\clubsuit\,\heartsuit$ & 0 \\
0 & 0 & 1 & $\heartsuit\,\clubsuit$ & 3,4 & $\clubsuit\,\heartsuit$ & 0 \\
0 & 1 & 0 & $\clubsuit\,\heartsuit$ & 5,6 & $\clubsuit\,\heartsuit$ & 0 \\
0 & 1 & 1 & $\heartsuit\,\clubsuit$ & 3,4 & $\clubsuit\,\heartsuit$ & 0 \\
1 & 0 & 0 & $\heartsuit\,\clubsuit$ & 3,4 & $\clubsuit\,\heartsuit$ & 0 \\
1 & 0 & 1 & $\clubsuit\,\heartsuit$ & 5,6 & $\clubsuit\,\heartsuit$ & 0 \\
1 & 1 & 0 & $\heartsuit\,\clubsuit$ & 3,4 & $\heartsuit\,\clubsuit$ & 1 \\
1 & 1 & 1 & $\clubsuit\,\heartsuit$ & 5,6 & $\heartsuit\,\clubsuit$ & 1 \\
\end{tabular}
\caption{Behavior of the AND protocol for all choices of
$(a,b,r) \in \{0,1\}^3$.}
\label{tab:and-correctness}
\end{table}

Here the output bit is obtained by decoding
$\clubsuit\,\heartsuit$ as $0$ and $\heartsuit\,\clubsuit$ as $1$.
As the table shows, for every $(a,b,r)$ the output bit is equal to
$a \land b$.
Therefore, the protocol correctly computes AND using six cards and one
random bisection cut.
\end{enumerate}

\subsubsection{XOR}\label{sec:xor}
We use an \emph{XOR primitive} that takes as input commitments to two
bits $a, b \in \{0,1\}$ and outputs a commitment to $a \oplus b$.
This primitive is realized by the four-card XOR protocol of Mizuki and Sone~\cite{MizukiSone2009}; in the version described below, it uses the two input commitments as its four cards, requires one random bisection cut, and uses no fresh helper cards.

\begin{enumerate}
\item \textbf{Arrange the initial commitments}\\
We prepare commitments $[a]$ and $[b]$ to the input bits
$a, b \in \{0,1\}$.
From left to right, we place the two-card encodings of $a$ and $b$ in
this order, forming a row of four cards:
\[
(c_1, c_2, c_3, c_4)
= (\text{first card of $[a]$}, \text{second card of $[a]$},
   \text{first card of $[b]$}, \text{second card of $[b]$}),
\]
so that $(c_1, c_2) = \mathrm{encode}(a)$ and
$(c_3, c_4) = \mathrm{encode}(b)$.

\item \textbf{Apply the permutation $\sigma = (2\ 3)$}\\
We apply the permutation $\sigma = (2\ 3)$ to the four cards, which
swaps positions $2$ and $3$ and leaves positions $1$ and $4$ unchanged.
In one-line notation, this permutation satisfies
$\sigma(1) = 1$, $\sigma(2) = 3$, $\sigma(3) = 2$, and $\sigma(4) = 4$,
and it transforms the row as
\[
(c_1, c_2, c_3, c_4)
\;\longrightarrow\;
(c_1, c_3, c_2, c_4).
\]

\item \textbf{Apply a random bisection cut (RBC)}\\
We split the four cards into a left block consisting of positions
$1$ and $2$ and a right block consisting of positions
$3$ and $4$, and apply one random bisection cut to these
two blocks.

Equivalently, we sample a uniformly random bit $r \in \{0,1\}$ and do
the following:
\begin{itemize}
\item if $r = 0$, we leave the order of the cards unchanged as
$(1,2,3,4)$;
\item if $r = 1$, we swap the left and right blocks, obtaining
$(3,4,1,2)$.
\end{itemize}
We again interpret the outcome of this RBC as implicitly generating the
uniform random bit $r$.

\item \textbf{Apply the inverse permutation $\sigma^{-1}$}\\
We apply the inverse permutation $\sigma^{-1} = (2\ 3)$ to the four
cards.
Since $\sigma$ is an involution, we have $\sigma^{-1} = \sigma$, so this
step again swaps positions $2$ and $3$.
After this step, the left two cards form a commitment to $a \oplus r$
and the right two cards form a commitment to $b \oplus r$, regardless
of the value of $r$.

\item \textbf{Reveal the first two cards and extract the XOR output}\\
We turn the first two cards (positions $1$ and $2$) face-up and reveal
the value of $a \oplus r$.
We then proceed by case analysis:
\begin{itemize}
\item If the revealed pattern is $\clubsuit\heartsuit$
(that is, $a \oplus r = 0$), then $r=a$ and the right two cards form
the commitment
$[b \oplus r] = [b \oplus a] = [a \oplus b].$
We therefore output the two cards at positions 3 and 4 as the
commitment to $[a \oplus b]$ and discard the two cards at positions
1 and 2.

\item If the revealed pattern is $\heartsuit\clubsuit$
(that is, $a \oplus r = 1$), then $r=\bar a$ and the right two cards form
the commitment
$[b \oplus r] = [b \oplus \bar a] = [\overline{a \oplus b}].$
By applying a single NOT operation, that is, by swapping the two cards
in the right pair, we obtain a commitment to $[a \oplus b]$. We then
discard the two cards at positions 1 and 2.

\end{itemize}
In either case, the only value that is revealed is $a \oplus r$, and
since $r$ is uniformly random, no information about the individual
values of $a$ and $b$ is leaked in an information-theoretic sense
\cite{MizukiSone2009}.
\end{enumerate}
\subsubsection{OR}\label{sec:or}
We use an \emph{OR primitive} that takes as input commitments to two
bits $a, b \in \{0,1\}$ and outputs a commitment to $a \lor b$.
By De Morgan’s law,
\[
a \lor b = \overline{\bar a \land \bar b}.
\]
so the OR primitive can be implemented by combining NOT operations on
commitments with the AND primitive~\cite{Mizuki2012}.
A NOT operation on a committed bit simply swaps the two cards in the
commitment and therefore requires no additional shuffle.
Consequently, our OR primitive can be realized with the same resources
as the AND primitive, namely six cards and one random bisection cut
\cite{Mizuki2012,Manabe2019}.

\subsubsection{Copy}\label{sec:copy}

Copying a committed bit is one of the central primitives in card-based
cryptography, and various copy protocols have been proposed using
different shuffle models and deck
assumptions~\cite{MizukiCopyRandomCut2021,MizukiPracticalCardCrypto2014,MizukiSone2009}.

In particular, Mizuki and Sone~\cite{MizukiSone2009} constructed a copy
protocol based on the random bisection cut (RBC).
Their construction works in full generality: given one input commitment
$[x]$ to a bit $x \in \{0,1\}$ and $k$ additional helper pairs, it
produces $k$ copies of $[x]$ simultaneously using a single RBC on
$2k+2$ cards in total.
We describe this protocol below.

\begin{enumerate}
\item \textbf{Arrange the initial commitments.}\\
We place the input commitment $[x]$ on the left and then arrange $k$
known commitments to $0$ to its right. For example, when $k=2$, the
initial arrangement is
\[
\begin{array}{rclrclrcl}
\qcard \qcard & \club \heart & \club \heart & \cdots &  & & & \\
\multicolumn{1}{c}{(x)} &  & 
 & &
\end{array}
\]
Thus, in general, we obtain a row of $2k+2$ cards
\[
(c_1,c_2,c_3,\ldots,c_{2k+1},c_{2k+2}),
\]
where $(c_1,c_2)$ is the commitment to $x$, and each pair
$(c_{2i+1},c_{2i+2})$ for $1 \le i \le k$ is the commitment
$\clubsuit\,\heartsuit$ (encoding $0$).

%We place the input commitment $[x]$ on the left and then arrange $k$
%known commitments to $0$ to its right:
%\[
%[x]\ \underbrace{[0]\ [0]\ \cdots\ [0]}_{k\ \text{pairs}}.
%\]
%Thus we obtain a row of $2k+2$ cards
%\[
%(c_1,c_2,c_3,\ldots,c_{2k+1},c_{2k+2}),
%\]
%where $(c_1,c_2)$ is the commitment to $x$, and each pair
%$(c_{2i+1},c_{2i+2})$ for $1 \le i \le k$ is the commitment
%$\clubsuit\,\heartsuit$ (encoding $0$).

\item \textbf{Gather odd and even positions.}\\
Apply a permutation that gathers the cards in odd positions to the left
half and those in even positions to the right half.
Specifically, the cards at positions $1,3,5,\ldots,2k+1$ are moved to
positions $1$ through $k+1$ (in this order), and the cards at positions
$2,4,6,\ldots,2k+2$ are moved to positions $k+2$ through $2k+2$ (in
this order).
After this step, the deck consists of a left block of $k+1$ cards and a
right block of $k+1$ cards.

\item \textbf{Apply a random bisection cut (RBC).}\\
We split the $2k+2$ cards into the left block (positions $1$ through
$k+1$) and the right block (positions $k+2$ through $2k+2$), and apply
one RBC.
This implicitly generates a uniformly random bit $r \in \{0,1\}$:
the blocks are left in place if $r=0$, and swapped if $r=1$.

\item \textbf{Redistribute to interleaved positions.}\\
We apply a second permutation that moves the left block to odd
positions and the right block to even positions.
Concretely, positions $1,\ldots,k+1$ are moved to positions
$1,3,5,\ldots,2k+1$, and positions $k+2,\ldots,2k+2$ are moved to
positions $2,4,6,\ldots,2k+2$.
After this step, the row consists of $k+1$ face-down pairs: the first
pair encodes $x \oplus r$, and each of the remaining $k$ pairs encodes
$r$.

\item \textbf{Reveal the first pair and obtain the copies.}\\
Turn the cards at positions $1$ and $2$ face-up to reveal $x \oplus r$.
\begin{itemize}
\item If they show $\clubsuit\,\heartsuit$ (encoding $0$), then
$r = x$, and the remaining $k$ pairs are all commitments to $[x]$.
We output them directly.
\item If they show $\heartsuit\,\clubsuit$ (encoding $1$), then
$r = \bar{x}$, and the remaining $k$ pairs are commitments to
$[\bar{x}]$.
We apply a NOT operation (swapping the two cards) to each pair to
obtain $k$ commitments to $[x]$.
\end{itemize}
In either case, the only value revealed is $x \oplus r$. Since $r$ is
uniformly random and independent of $x$, no information about $x$ is
leaked in the information-theoretic
sense~\cite{MizukiSone2009}.
\end{enumerate}
In this primitive, the input commitment is consumed because the first pair is revealed. 
When we say that we create a copy of an existing commitment while preserving the original commitment in a later protocol, we run the above primitive with $k=2$, return one of the two output commitments to the original position, and use the other as the copy.  Thus, one preserved copy requires four helper cards in addition to the original commitment.

\subsection{Batching technique}\label{sec:batching}
When protocols use the pile-scramble shuffle, it is possible to batch
several pairwise disjoint pile-scramble shuffles into a single
pile-scramble shuffle with the help of a few additional cards.
This idea, known as the batching technique, was introduced by
Shinagawa and Nuida~\cite{ShinagawaNuida2021}.

Informally, suppose we have several pile-scramble shuffles
$S_1, S_2, \ldots$ that act on disjoint collections of piles,
so that they could in principle be executed in parallel.
The batching technique lets us perform all of these shuffles using a
single physical pile-scramble operation.

The key idea is to attach a short identifier to each pile using
helping cards, indicating to which shuffle $S^{(t)}$ it belongs.
We then place all piles together and apply one global pile-scramble
shuffle.
After the shuffle, we reveal only the identifiers (keeping the payload
cards face-down) and use them to regroup the piles according to their
original shuffle $S^{(t)}$.
In this way, a single physical pile-scramble shuffle implements all
of $S_1, S_2, \ldots$ simultaneously, at the cost of
introducing a small number of helping cards.

As an example, consider the $n$-input AND protocol of Kuzuma
et al.~\cite{Kuzuma2022SingleShuffle}, which uses shuffles
$S_1, S_2, \ldots, $ $S_{n-1}$.
Among these, the shuffles with odd indices
$S_1, S_3, S_5, \ldots$ can be batched into a single pile-scramble
shuffle, and likewise the shuffles with even indices
$S_2, S_4, S_6, \ldots$ can be batched into another one.
Thus, by using batching, the AND protocol can be implemented using
only two pile-scramble shuffles, at the cost of introducing a few
extra helping cards~\cite{Kuzuma2022SingleShuffle,ShinagawaNuida2021}.

We note that Kuzuma et al. also describe how to combine multiple
shuffles into a single shuffle at a purely mathematical level so that
only one shuffle is applied.
However, the resulting shuffle is no longer a pile-scramble and is
difficult to implement physically~\cite{Kuzuma2022SingleShuffle}.
In this work, we exclusively use batching techniques that can be
realized by physically feasible pile-scramble shuffles.
\section{Moderatorless Werewolf}\label{sec:werewolf}

The party game \WW (also known as Mafia) is a social deduction game
in which a small informed minority gradually eliminates a larger
uninformed majority by deception and by steering the voting process.
In this section, we describe a moderatorless implementation of a
standard variant of \WW using our card-based framework.
More precisely, we identify the tasks that a generic moderator must
perform and explain how our protocol implements them using only
publicly observable card operations.
The rule set considered in this section is based on common versions of
Werewolf, including \emph{The Werewolves of Miller's Hollow}~\cite{WerewolvesMillersHollowRules}, \emph{Are You a Werewolf?}~\cite{AreYouAWerewolfRules},
and \emph{Ultimate Werewolf}~\cite{UltimateWerewolfRules}.  These versions differ in some details, such as the available roles, the night-time procedures, and the treatment of special events.
  Our protocol is not tied to any particular rule set: such differences can be accommodated by minor modifications of the role-processing cards and of the order in which the corresponding subprotocols are executed.
We organize the description according to the three phases of the
game: the pre-game phase, the day phase, and the night phase.
In this variant, every player belongs to exactly one of two sides:
the \emph{village} side, consisting of Villagers, the Seer, and the
Bodyguard (Doctor), or the \emph{werewolf} side, consisting of one or more
Werewolves.
The village side wins if all Werewolves are eliminated, and the
werewolf side wins as soon as the number of surviving Werewolves is
at least as large as the number of surviving non-Werewolf players.

\subsection{High-level game flow}\label{subsec:ww-flow}

In this subsection, we outline the overall flow of a standard
variant of \WW and the corresponding tasks that a moderator would
normally perform.
These are exactly the points that our moderatorless protocol must
realize using card-based procedures.
The game begins with a pre-game phase and then proceeds by repeating
day and night phases.
In each phase, the players execute the appropriate card-based
subprotocols.

\paragraph{Pre-game phase.}
At the beginning of the game, the moderator randomly chooses a role
assignment consistent with the publicly known multiset of roles and
privately informs each player of their role.
Only the moderator knows the complete role assignment.
At this stage, the Werewolves are the only players who learn any
other players' roles: each Werewolf learns which other players are
Werewolves.
In a digital implementation, this can be achieved via private
channels within an application.
In an analog implementation, it is typically realized by instructing
all players to close their eyes and then asking only the Werewolves
to open their eyes and silently identify one another.
At this point, the publicly known multiset of roles and the total
number of players are common knowledge among all players.
Apart from their own roles, non-Werewolf players receive no
additional information about the role assignment.
\begin{comment}
At the beginning of the game, the moderator samples a role
assignment consistent with a publicly known multiset of roles and
privately communicates to each player her own role.
Only the moderator has complete information about all roles.
The only players who learn other players' roles at this stage are
the Werewolves: each Werewolf learns the identities of all other
Werewolves.
In a digital implementation, this can be achieved via private
channels within an application; in an analog implementation, it is
typically realized by instructing all players to close their eyes,
then asking only the Werewolves to open their eyes and silently
identify one another (and similarly for any other role with shared
information, such as Masons).
For all other players, the only facts that are common knowledge at
this point are the public multiset of roles and the total number of
players; the Werewolves also know these public parameters.
\end{comment}

\paragraph{Day phase.}
During each day phase, the moderator intervenes only after the
public discussion, in the voting phase.
After the discussion ends, the moderator collects one vote from each
player, aggregates the votes, and identifies the candidate with the
largest number of votes.
The result is then publicly announced to all players.
%In some variants, voting is carried out publicly, but in others it is required to be anonymous; accordingly, our protocol explicitly supports both public and anonymous voting, for example by providing individual input channels for anonymous ballots. The player with the highest number of votes is \emph{executed} and removed from the game.

\paragraph{Night phase.}
During each night phase, the moderator must sequentially collect
private actions from the roles that act at night, typically in the
order Seer, Werewolves, and Bodyguard.
In the Seer step, if the Seer is alive, she privately selects a
target player, and the moderator privately tells her whether the
target is a Werewolf.
In the Werewolf step, the Werewolves---who have full information
about each other's identities---jointly and secretly select a single
target player to attack.
In the Bodyguard step, if the Bodyguard is alive, she privately selects
one player to protect; if that player is attacked by the Werewolves
in the same night, the attack is canceled and the player does not
die.
The moderator takes the inputs from the Werewolves and the Bodyguard
and, following a fixed priority rule (for example, successful
protection cancels the corresponding attack), determines whether any
deaths occur during the night and, if so, which players die.
The resulting death information is carried over to the next day
phase and publicly announced at the beginning of that phase.

\subsection{Used cards}\label{subsec:used-cards}

We consider a game of \WW with $n$ players.
There are four roles: Villager, Werewolf, Seer, and Bodyguard, and
their numbers are fixed in advance to $n-k-2$, $k$, $1$, and $1$,
respectively.
At the beginning of the game, each player receives two bundles of
cards: a bundle of \emph{player-number cards} and a bundle of
\emph{role-related cards}.
Each player is assigned a distinct player number from $1$ to $n$.
For player~$i$, the player-number card is the numeric card whose
front side shows the integer $i$.
These cards are used to specify target players in the protocols.
The role-related cards serve both to inform each player of her own
role and to enable her to perform the corresponding role-specific
procedures in the game.
In other words, the assignment of roles is realized by randomly
distributing the bundles of role-related cards.
Among the role-related cards, we call those used in the role-specific
procedures \emph{role-processing cards}.
Each role-processing card bundle consists of an ordered sequence of
suit cards \club~and \heart, and both the multiset and the order of
these cards depend on the player's role.
The order within a bundle is part of its encoding.
When a bundle is used as input to one of the role-specific
subprotocols described later, it is handed over as a bundle in this
prescribed order, and players may not rearrange its cards except as
specified by the public procedure.
These cards are placed and manipulated according to publicly
specified procedures in later sections, but the way they are prepared
and distributed depends only on the underlying role assignment. 
%
\begin{comment}
We consider a game of \WW with $n$ players.
There are four roles: Villager, Werewolf, Seer, and Bodyguard, and
their numbers are fixed in advance to $n-k-2$, $k$, $1$, and $1$,
respectively.
At the beginning of the game, each player receives two bundles of
cards: a bundle of \emph{player-number cards} and a bundle of
\emph{role-related cards}.

Each player is assigned a distinct player number from $1$ to $n$.
For player~$i$, the player-number card is the numeric card whose
front side shows the integer $i$.
These cards are used to specify target players in the protocols.

The role-related cards serve both to inform each player of her own
role and to enable her to perform the corresponding role-specific
procedures in the game.
In other words, the assignment of roles is realized by randomly
distributing the bundles of role-related cards.
Among the role-related cards, we call those used in the role-specific
procedures \emph{role-processing cards}.
Each role-processing card bundle consists of a number of suit cards
\club and \heart, and the multiset and arrangement of these cards
depend on the player's role.
These cards are placed and manipulated according to publicly
specified procedures in later sections, but the way they are prepared
and distributed depends only on the underlying role assignment.
\end{comment}
%
\subsection{Protocols by purpose}\label{subsec:purpose-protocols}

In this subsection, we classify the card-based subprotocols that we
use in our moderatorless implementation of \WW according to their
purpose.
Each class collects protocols that share the same high-level task and
can be instantiated by a common design pattern.
We distinguish the following three classes of protocols:
\begin{enumerate}[label=(\roman*)]
  \item \textbf{Role-internal information-sharing protocols.}
  A role-internal information-sharing protocol allows players with a
  specific role to share a short piece of information among
  themselves, while players with any other role learn nothing about
  that information.
  In this paper, the \emph{werewolf introduction protocol} and the
  \emph{attack-target discussion protocol} belong to this class.
  In the werewolf introduction protocol, Werewolf players share their
  own player numbers so that they can learn who the other Werewolves
  are.
  In the attack-target discussion protocol, Werewolf players share
  the player numbers of the players they wish to attack, enabling
  limited coordination of their intended targets.
  \item \textbf{Action-specification protocols.}
  In an action-specification protocol, at most one player with role~$Y$
  can secretly choose at most one player number $x$, different from
  their own player number, as the target of an action.
  Until the action is actually executed, no other player learns who
  chose which target.
  If the action on $x$ is eventually executed and succeeds, only the
  result is revealed, without revealing who chose $x$.
  If no target is chosen, the action is applied only to dummy cards.
  If different types of actions can be applied to the same player
  $x$, the outcome (success or failure) is assumed to be determined
  by a rule fixed in advance.
  If the action is executed but fails to produce a public effect, the
  protocol reveals neither the chosen target~$x$ nor who chose it.
  In this paper, we treat it as cheating if more than one player with
  role~$Y$ chooses a target at the same time.
  The werewolf attack protocol and the bodyguard protection protocol
  belong to this class.
  \item \textbf{Attribute-judgment protocols.}
  In an attribute-judgment protocol, each role is associated with a
  binary value encoded by a pair of suit cards, either
  \(\heart\club = 1\) or \(\club\heart = 0\).
  A player with a designated judging role may secretly select a
  player and learn the corresponding binary value of that player's
  role.
  All other players, however, remain ignorant of both the selected
  player and the obtained value.
  The seer-judgment protocol and the medium protocol (for extended
  rules) belong to this class.
  In the seer-judgment protocol, Werewolf roles are encoded as~1 and
  all other roles as~0, and only the Seer learns the resulting bit.
\end{enumerate}
In the following protocols, we use separate parameters for the sizes of
the relevant card bundles.  Let $\hshr$ denote the number of
cards in one sharing-information-card bundle, and let $\hact$
denote the number of cards in one action-processing-card bundle.
%, and $\hatt$ the number of cards in one role-value-card bundle used for attribute judgment.

\begin{comment}
  \item \textbf{Action-specification protocols.}
  In an action-specification protocol, each player with a designated
  role $Y$ may secretly specify one player number $x$ as the target
  of an action.
  Until the action is actually executed, no other player learns who
  specified whom.
  If the action on $x$ is eventually executed and succeeds, only the
  result is revealed, without revealing who specified $x$.
  If different types of actions can be applied to the same player
  $x$, the outcome (success or failure) is assumed to be determined
  by a rule fixed in advance.
  If the execution fails, the identity of~$x$ is not revealed.
  In this paper, we treat it as cheating if more than one player with
  role~$Y$ specifies a target at the same time.
  The werewolf attack protocol and the bodyguard protection protocol
  belong to this class.
\end{comment}

\subsubsection{Role-internal information-sharing protocol}
\label{subsubsec:role-sharing}
In this subsection, we describe the role-internal information-sharing
protocol.
Let $n'$ denote the current number of surviving players.  
Here and below, we index the surviving players as $1,\ldots,n'$ only for notational
convenience; their original player-number cards are not reassigned.
At the beginning of the protocol, each player is classified, according to her role, into either a \emph{sharing player}, who shares information with other players of the same role, or a
\emph{non-sharing player}, who does not share information.

Each sharing player wishes to share a short piece of information
with all other sharing players, while non-sharing players must learn
nothing about this information.
As dummy cards, we use numeric cards with value~$0$ printed on their
faces.
Each player holds the following two types of role-processing card
bundles.

\begin{enumerate}
  \item \textbf{Sharer-identification cards.}
  Each sharing player has a bundle of suit cards \(\heart\club\),
  whereas each non-sharing player has a bundle \(\club\heart\).
  We use these bundles as sharer-identification cards.

  \item \textbf{Sharing-information cards.}
  Recall that $\hshr$ denotes the number of cards needed to represent one unit of
  information to be shared; the value of $\hshr$ is known to all players.
  Each sharing player has $n'-1$ bundles of $\hshr$ cards, all encoding
  the information she wants to share.
  Each non-sharing player has $n'-1$ bundles of $\hshr$ dummy cards.
  We call each of these bundles a sharing-information-card bundle.
  For each player, the $n'-1$ sharing-information-card bundles she
  holds are all identical.
\end{enumerate}

The protocol proceeds as follows.

\begin{enumerate}[label=\arabic*.]
  \item Arrange face-down card bundles in a grid with $2n'$ rows and
$n'+1$ columns, and denote the bundle in row~$i$ and column~$j$ by
$p_{i,j}$.
  The bundles are placed so that the following conditions hold.
  \begin{itemize}
    \item For each $1 \le r \le n'$ and each $1 \le j \le n'-1$,
player~$r$ places one of her own sharing-information-card bundles
on $p_{2r-1,j}$.
She also places her sharer-identification-card bundle on
$p_{2r-1,n'}$ and her player-number card on $p_{2r-1,n'+1}$.
    \item For each $1 \le r \le n'$ and each $1 \le j \le n'-1$, a
bundle of $\hshr$ dummy cards is placed on each $p_{2r,j}$.
In addition, on each $p_{2r,n'}$ we place a commitment
\(\heart\club\), and on each $p_{2r,n'+1}$ we place a player-number
card with value~$0$.
  \end{itemize}

  \item For each $1 \le r \le n'$, we create a copy
\(p^{\ast}_{2r-1,n'}\) of the commitment on~$p_{2r-1,n'}$ using the
copy primitive, and compute the XOR of \(p^{\ast}_{2r-1,n'}\) and
$p_{2r,n'}$ to obtain a new commitment $q_r$.
We then replace the bundle on $p_{2r,n'}$ with $q_r$.

  \item We gather each row into a single pile and apply a
  pile-scramble shuffle to the $2n'$ piles.

    \item We publicly reveal only the $n'$-th column.
    Let \(L = (l_1,l_2,\dots,l_{n'})\) be the ordered list of row indices,
    in increasing order,  
    in which the revealed bundle is
\(\heart\club\).
  We cyclically permute the sharing-information-card bundles among
  these rows so that each selected row receives one
  sharing-information-card bundle from each of the other selected rows.
  More precisely, let \(p'_{i,j}\) denote the bundle in row~$i$ and
  column~$j$ after this permutation.  For \(1 \le t \le n'\) and
  \(1 \le j \le n'-1\), define
  \[
    s(t,j) = 1 + ((t+j-1) \bmod n').
  \]
  Then, we rearrange the bundles so that
  \[
    p'_{l_t,j} = p_{l_{s(t,j)},j}
      \qquad
      (1 \le t \le n',\ 1 \le j \le n'-1),
  \]
  while all other bundles remain in their original positions.

\item We turn the sharer-identification-card bundles in the $n'$-th
column face-down again.  We then gather each row into a single pile
and apply a pile-scramble shuffle to the $2n'$ piles.

  \item We turn face-up the player-number cards in column~$n'+1$.
  For each row whose player-number card shows a value
  \(i \in \{1,\dots,n'\}\), we hand the entire row of bundles to
  player~$i$.
  Each player privately inspects the bundles she receives.
  Rows whose player-number card has value~$0$, as well as all
  bundles that have been inspected, are discarded face-down.
\end{enumerate}

We now argue that the above protocol works correctly.
It is enough to show the following three properties.
\begin{enumerate}
  \item A non-sharing player obtains no information contained in the
  sharing-information-card bundles placed by other surviving players.
  \item Each sharing player obtains, for every other sharing player,
  at least one sharing-information-card bundle placed by that player.
  \item A non-sharing player cannot learn the number of sharing
  players.
\end{enumerate}

We first prove the first property.
In Steps~3 and~5, each row is gathered into a single pile and a
pile-scramble shuffle is applied to the rows.
Thus these operations permute rows but do not change the contents
within any row.
In Step~4, sharing-information-card bundles are permuted only among
the rows whose sharer-identification card in column~$n'$ is revealed
as \(\heart\club\).
For a non-sharing player, the row containing their player-number card is not one of these selected rows, because the selected row associated with that player is the dummy row.
Therefore, the contents of the row eventually returned to a
non-sharing player are unchanged.
Consequently, when a non-sharing player privately inspects the
bundles received in Step~6, the sharing-information-card bundles they
see are only their own dummy bundles; hence they obtain no
information from the sharing-information-card bundles of other
surviving players.

We next prove the second property.
Consider two distinct sharing players with player numbers \(a\) and
\(b\).
After Step~3, let \(l_\alpha\) and \(l_\beta\) be the selected rows
that contain the player-number cards of players \(a\) and \(b\),
respectively; that is, \(l_\alpha\) and \(l_\beta\) are the
\(\alpha\)-th and \(\beta\)-th elements of the ordered list
\(L=(l_1,\dots,l_{n'})\).
Since both players are sharing players, both of these rows are
selected rows.
By the cyclic permutation in Step~4, for the target row
\(l_\alpha\), the bundles placed in columns \(1,\dots,n'-1\) are taken from the selected rows other than \(l_\alpha\), exactly
once each.
In particular, there exists some \(j \in \{1,\dots,n'-1\}\) such that
\[
  p'_{l_\alpha,j} = p_{l_\beta,j}.
\]
Thus player \(a\) receives at least one sharing-information-card
bundle originally placed by player \(b\).
The same argument with \(a\) and \(b\) interchanged shows that player
\(b\) receives at least one such bundle from player \(a\).
Since the pair of sharing players was arbitrary, every sharing player
receives at least one sharing-information-card bundle from every other
sharing player.

Finally, we prove the third property.
By Step~2, for each player, exactly one of the two corresponding
rows has sharer-identification value \(\heart\club\) in column~$n'$,
and the other has value \(\club\heart\).
Therefore, after Step~2, column~$n'$ contains exactly \(n'\) bundles of
type \(\heart\club\) and exactly \(n'\) bundles of type
\(\club\heart\), regardless of how many players are sharing players.
Hence the cards revealed in Step~4 do not reveal the number of
sharing players.
Moreover, by the first property, a non-sharing player receives no
sharing-information-card bundle placed by another player in Step~6.
Thus a non-sharing player obtains no additional information from
which the number of sharing players could be inferred.

Therefore, the protocol satisfies the required correctness and privacy
properties.

\subsubsection{Action-Specification Protocol}
This subsection describes the action-specification protocol.
Let $n'$ be the number of surviving players at the current point.
Initially, according to their roles, the players are classified into two types: players who can choose at most one player other than themselves as an action target (\emph{action-specifying players}) and the remaining players (\emph{non-action-specifying players}).
An action-specifying player has one $\heart \club$ bundle and $n'-1$ $\club \heart$ bundles as role-processing cards, while a non-action-specifying player has $n'$ $\club \heart$ bundles as role-processing cards.
Thus, both types of players have $n'$ role-processing-card bundles.
Recall that $\hact$ is the number of cards required to process the action.
We assume that $\hact$ is known to all players.
Each player has a predetermined bundle of $\hact$ cards for processing the action as their \emph{action-processing cards}.
Note that a player does not necessarily know the contents of their own action-processing cards.
In the action-specification protocol, an action-specifying player chooses at most one player other than themselves.
If a player is chosen, the predetermined action, such as revealing cards or replacing cards with dummy cards, is applied to the action-processing cards of the chosen player.
If no player is chosen, the action is applied only to dummy cards.
We assume that the action to be applied to the action-processing cards is known in advance to all players.
The protocol is as follows.

\begin{enumerate}
    \item Place face-down card bundles in a grid with $n'+1$ rows and $n'+1$ columns satisfying the following conditions. Let $p_{i,j}$ denote the card bundle in row $i$ and column $j$.
    \begin{itemize}
        \item For each $1\leq j\leq n'$, place a $\heart \club$ bundle at $p_{1,j}$; this row represents the case in which player $j$ chooses no target. Place a bundle of $\hact$ dummy cards at $p_{1,n'+1}$.
        \item For each $1\leq j\leq n'$, the position $p_{j+1,j}$ is the self-target position of player $j$. We place a fixed $\club \heart$ bundle at this position.
        \item For each $2\leq i\leq n'+1$ and $1\leq j\leq n'$ with $i\neq j+1$, player $j$ places one of their role-processing-card bundles at $p_{i,j}$. Each player leaves one role-processing-card bundle unused; unused bundles are kept face down and are not used in the protocol.
        \item For each $2\leq i\leq n'+1$, place the action-processing cards of player $i-1$ at $p_{i,n'+1}$.
    \end{itemize}

    \item For each $2\leq i\leq n'+1$ and $1\leq j\leq n'$, obtain a copy $p^{*}_{i,j}$ of the bundle $p_{i,j}$ using the copy protocol. For each $1\leq j \leq n'$, compute a commitment $p_{j,\mathrm{or}}$ to the OR of the commitments $p^{*}_{2,j},p^{*}_{3,j},\dots,p^{*}_{n'+1,j}$. Then compute a commitment to the XOR of $p_{1,j}$ and $p_{j,\mathrm{or}}$, and replace $p_{1,j}$ with the resulting commitment.

    \item We gather each row into a single pile and apply a
    pile-scramble shuffle to the $n'+1$ piles.

    \item We gather each of the first $n'$ columns into a single pile and
    apply a pile-scramble shuffle to these $n'$ piles.

    \item Repeat the following operation $n'$ times. In the $i$-th operation, reveal the cards in the $i$-th column. If the card in the $j$-th row of that column is a $\heart \club$ bundle, execute the action on the bundle at $p_{j,n'+1}$. Then turn all revealed cards face down again, gather each row into a single pile, and apply a pile-scramble shuffle to the $n'+1$ piles.
\end{enumerate}

If we additionally wish to verify the number of players who chose a player other than themselves, we perform the following procedure after Step~1.
\begin{enumerate}
    \item For each $2\leq i\leq n'+1$ and $1\leq j\leq n'$, obtain a copy of the bundle $p_{i,j}$ using the copy protocol.
    \item Regard each obtained copied bundle as an individual bundle and perform a pile-scramble shuffle.
    \item Reveal all face-down cards and count the number of $\heart\club$ card pairs. This number is equal to the number of players who chose a player other than themselves.
\end{enumerate}

If we wish to make it possible to identify the original owner of each action-processing-card bundle after the action, we perform the following procedure.
\begin{enumerate}
    \item Extend the arrangement in Step~1 to $n'+2$ columns and $n'+1$ rows. Place a player-number card with value $0$ at $p_{1,n'+2}$. For each $2\leq i \leq n'+1$, place the player-number card with value $i-1$ at $p_{i,n'+2}$.
    \item Carry out Steps~2 through~5 of the action-specification protocol in the usual way, including column $n'+2$ whenever rows are gathered into piles. After the action-specification protocol terminates, gather each row into a single pile and apply a pile-scramble shuffle to the $n'+1$ piles.
    \item Reveal the cards in column $n'+2$. For $x\neq 0$, if the player-number card with value $x$ appears in row $i$, then the bundle at $p_{i,n'+1}$ is the action-processing-card bundle of player $x$. If the card with value $0$ appears in row $i$, then the bundle at $p_{i,n'+1}$ is the dummy bundle.
\end{enumerate}

We now explain why the above protocol works correctly.
Since all role-processing cards distributed to a non-action-specifying player are $\club \heart$ bundles, every role-processing-card bundle placed by such a player is a $\club \heart$ bundle.
When an action-specifying player places their bundles in Step~1, they can do one of the following:
\begin{itemize}
    \item place $\club \heart$ bundles in all selectable positions of their column;
    \item place a $\heart \club$ bundle in exactly one selectable position of their column and place $\club \heart$ bundles in all other selectable positions of that column.
\end{itemize}
Here, the selectable positions of column $j$ are the positions $p_{i,j}$ with $2\leq i\leq n'+1$ and $i\neq j+1$.
The self-target position $p_{j+1,j}$ is always a fixed $\club \heart$ bundle.
In either case, outside the first row, each of the first $n'$ columns contains at most one $\heart \club$ bundle.
By Step~2, if a $\heart \club$ bundle is placed in column $j$ at a selectable position, then $p_{1,j}$ becomes a $\club \heart$ bundle; otherwise, it remains a $\heart \club$ bundle.
Therefore, after Step~2 is completed, each of the first $n'$ columns contains exactly one $\heart \club$ bundle, satisfying the following conditions:
\begin{itemize}
    \item if no $\heart \club$ bundle was placed in column $j$ at a selectable position in Step~1, then $p_{1,j}$ is the $\heart \club$ bundle;
    \item if a $\heart \club$ bundle was placed at a selectable position $p_{i,j}$ in Step~1, then $p_{i,j}$ is the $\heart \club$ bundle.
\end{itemize}

Define a function $\rho$ as follows.
Given an input $j$, let $i$ be the row containing the $\heart \club$ bundle in column $j$ after Step~2.
Then $\rho(j)$ returns the bundle at $p_{i,n'+1}$, which is either an action-processing-card bundle or the dummy bundle in the first row.
After the shuffle in Step~3, if the card in row $i'$ and column $j$ is a $\heart \club$ bundle, then $p_{i',n'+1}$ is guaranteed to be the bundle $\rho(j)$.
Thus, in Step~5, the action is correctly executed on the bundle in column $n'+1$ of the row in which a $\heart \club$ bundle is revealed.
Moreover, for an original column $j$, the self-target position $p_{j+1,j}$ is fixed to be a $\club \heart$ bundle.
Thus, before the row shuffle in Step~3, column $j$ has no $\heart \club$ bundle in the row containing player $j$'s own action-processing-card bundle.
Step~3 only permutes whole rows, Step~4 only permutes whole columns, and the row shuffle after each iteration of Step~5 also permutes whole rows, so this property is preserved until the column is revealed.
Therefore, the action can never be applied to the action-processing-card bundle of the player who made the choice.
Furthermore, since a pile-scramble shuffle over the rows is performed in Step~3
and after each column operation in Step~5, no information can be learned from dependencies between the operations on different columns.
Since each of the first $n'$ columns is guaranteed to contain exactly one $\heart \club$ bundle, the revealed column of role-processing cards also leaks no information to the players.

Next, we show that the number of players who chose a player can be correctly detected.
At the end of Step~1, among the positions in rows $2,\dots,n'+1$, a column corresponding to a player who chose a player contains exactly one $\heart \club$ bundle, while all other such columns contain none.
Therefore, the number of such players can be determined by counting the number of $\heart \club$ bundles among the copied bundles. Moreover, because a pile-scramble shuffle is performed before counting the number of $\heart \club$ bundles, it is impossible to learn which player made each choice.

Finally, we explain why the original owner of each action-processing-card bundle can be identified after the action.
The extra column $n'+2$ contains player-number cards, and each such card is initially placed in the same row as the corresponding player's action-processing-card bundle in column $n'+1$.
During the action-specification protocol, these two columns are never separated: whenever the rows are shuffled, each row is gathered into a single pile, so the player-number card in column $n'+2$ moves together with the action-processing-card bundle in column $n'+1$ of the same row.
The action itself may modify the action-processing-card bundle, but it does not change its associated player-number card.
Therefore, after column $n'+2$ is revealed, the player-number card in each row identifies the original owner of the action-processing-card bundle in column $n'+1$ of that same row.
In particular, for $x\neq 0$, if the card labeled $x$ appears in row $i$, then $p_{i,n'+1}$ is the bundle originally owned by player $x$.
If the card labeled $0$ appears in row $i$, then $p_{i,n'+1}$ is the dummy bundle.

\subsubsection{Attribute-judgment protocol}\label{sec:Attribute-judgment protocol}
This subsection describes the attribute-judgment protocol.
Let $n'$ be the number of surviving players at the current point.
Initially, according to their roles, the players are classified into two types: players who can choose a player and learn that player's attribute (\emph{attribute-judging players}) and the remaining players (\emph{non-attribute-judging players}).
In this protocol, an attribute-judging player chooses one player and learns the encoded value corresponding to the chosen player's role.
Each player holds the following two types of role-processing-card bundles.

\begin{enumerate}
    \item \textbf{Role-value cards.}
    Each player has $n'$ bundles encoding the value associated with their role.
    Depending on the role, these bundles are either $\heart \club$ bundles or $\club \heart$ bundles.

    \item \textbf{Attribute-judgment cards.}
    The $\club\club$ bundles used here are selector bundles and are
not bit commitments.
    An attribute-judging player has one $\heart \club$ bundle and $n'-1$ $\club \club$ bundles.
    A non-attribute-judging player has $n'$ $\club \club$ bundles.
\end{enumerate}

The protocol is as follows.
\begin{enumerate}
    \item Repeat Steps~2 through~4 for each $i=1,2,\dots,n'$.
    \item %In the $i$-th iteration, player $i$ places their role-value-card bundles face down in two vertical columns, one card per column. 
In the $i$-th iteration, player $i$ places their $n'$ role-value-card bundles face down in two vertical columns, one bundle in each row.
Each role-value-card bundle consists of two suit cards, and its first and second cards are placed in the two role-value columns of the corresponding row.
    %Then, for each $1\leq j \leq n'$, the player with player number $j$ places one of their attribute-judgment-card bundles in two vertical columns immediately to the left of the role-value cards that player $i$ placed in row $j$.
    Then, for each $1 \le j \le n'$, player $j$ takes one attribute-judgment-card bundle and places its two cards in the two vertical columns immediately to the left of the role-value cards placed in row $j$.
    \item Gather each of the four vertical columns into a single pile, and apply a pile-shifting shuffle to these four piles.
    \item After the shuffle, for each $1\leq j \leq n'$, hand the four cards now in row $j$ to player $j$. Each player privately views the cards they receive. Afterward, the cards are turned face down again and discarded.

\end{enumerate}

We now explain why the above protocol works correctly.
Since each iteration is independent, it suffices to show that the $i$-th iteration works correctly.
Player $i$ never sees the cards placed by the other players, and therefore learns no information.
For any $j\neq i$, player $j$ only sees the cards in row $j$, and therefore does not learn the contents of the attribute-judgment cards placed by the other players in Step~2.
Suppose that a player places a $\club \club$ bundle as their attribute-judgment card.
Then, regardless of the value of the role-value card placed by player $i$, the four cards received by that player in Step~4 consist of three $\club$ cards and one $\heart$ card.
Thus, this player learns no information about player $i$'s role value.
On the other hand, suppose that a player places a $\heart \club$ bundle as their attribute-judgment card.
Then the order of the cards received in Step~4 allows this player to identify player $i$'s role-value card.
Specifically, if the two $\heart$ cards are adjacent when the received sequence is regarded cyclically, then the role-value card is $\club \heart$; otherwise, it is $\heart \club$.
Since each attribute-judging player has only one $\heart \club$ bundle, they can choose exactly one player and learn that player's encoded role value.

\subsection{Implementing the phases of \WW}
This subsection explains how to implement the pre-game phase, the day and
night phases, and the victory-check phase using the above purpose-specific
protocols and basic primitives.

\paragraph{Pre-game phase.}
In the pre-game phase, we perform the following procedure.
\begin{itemize}
    \item \emph{Werewolf introduction protocol.} The Werewolf players share their player numbers with the other Werewolf players.
\end{itemize}
The Werewolf introduction protocol can be implemented using the role-internal information-sharing protocol.
Specifically, the players whose role is Werewolf are treated as sharing players, and all other players are treated as non-sharing players.
The shared-information card held by each sharing player is their own player-number card.
Thus, by using the role-internal information-sharing protocol, each Werewolf player can share their player-number card with the other Werewolf players.

\paragraph{Day phase.}
During the day phase, the players discuss and then vote to execute one player.
The player with the most votes is executed.
This step requires no special card protocol and can be carried out by public declarations or by pointing.

\paragraph{Night phase.}
In the night phase, we perform the following procedures.
\begin{itemize}
    \item \emph{Attack-target sharing protocol.} The Werewolves exchange information with the other surviving Werewolf players and decide tonight's attack target.
    \item \emph{Seer protocol.} The Seer chooses one player and obtains a binary encoded value indicating whether the chosen player is a Werewolf.
    \item \emph{Bodyguard protocol.} The Bodyguard chooses one player other than themselves and protects the chosen player from the Werewolf attack by replacing that player's action-processing-card bundle with a dummy card.
    \item \emph{Attack protocol.} One representative Werewolf chooses one player other than themselves and attacks the chosen player. We do not consider the case in which a Werewolf attacks another Werewolf.

\end{itemize}

The attack-target sharing protocol can be implemented by repeatedly using the role-internal information-sharing protocol.
The players whose role is Werewolf are treated as sharing players, and all other players are treated as non-sharing players.
The shared-information cards are implemented using player-number cards and $\club,\heart$ cards.
We assume that all players agree in advance on what information is sent to determine the attack target and on how many times the role-internal information-sharing protocol is repeated.

The Seer protocol can be implemented using the attribute-judgment protocol.
The player whose role is Seer is treated as an attribute-judging player, and all other players are treated as non-attribute-judging players.
The role-value cards of Werewolf players are $\heart \club$, and the role-value cards of all other players are $\club \heart$.

The Bodyguard protocol can be implemented using the action-specification protocol.
The player whose role is Bodyguard is treated as an action-specifying player, and all other players are treated as non-action-specifying players.
Each player's action-processing-card bundle is a one-card bundle containing their own player-number card.
The action is to replace the chosen card with a dummy card, namely a card labeled $0$.

The attack protocol can be implemented using the action-specification protocol.
In this paper, we do not consider attacks against other Werewolves.
The players whose role is Werewolf are treated as action-specifying players, but exactly one of them is supposed to choose a target; the others choose no target.
For the attack protocol, each player's action-processing-card bundle is the bundle obtained after the Bodyguard protocol has been applied.
The action is to reveal the chosen card.
If the revealed card is not a dummy card, then the player corresponding to the revealed player-number card is regarded as having been attacked.
Moreover, to check whether two or more Werewolf players chose a target, we use the procedure for verifying the number of players who chose a player other than themselves in the action-specification protocol.

\paragraph{Victory-check phase.}
The victory condition is checked whenever a player dies.
In the victory-check phase, we check the following two conditions.
\begin{itemize}
    \item whether the number of Werewolf players is at least the number of non-Werewolf players;
    \item \emph{Werewolf-survival judgment protocol:} whether at least one player whose role is Werewolf is still alive.
\end{itemize}
%No protocol is needed to determine whether the number of Werewolf players is at least the number of non-Werewolf players, because the Werewolf players can determine this themselves.
The Werewolf-side winning condition is self-detectable by the Werewolves: once the number of surviving Werewolves is at least the number of surviving non-Werewolf players, the Werewolves can reveal themselves, and the condition becomes publicly verifiable.  Thus, no additional card-based protocol is needed for this condition.

In contrast, the village-side winning condition, namely that no Werewolf survives, cannot be certified by a voluntary declaration of the Werewolf side.  We therefore use the Werewolf-survival judgment protocol to test whether at least one Werewolf remains alive.

In the Werewolf-survival judgment protocol, each player whose role is Werewolf has $\heart \club$ as their role-processing card, and each other player has $\club \heart$ as their role-processing card.
By applying the OR primitive to all players, we can determine whether at least one Werewolf remains alive.

Combining these subprotocols, we obtain a card-based implementation of the
moderator's required tasks in this variant of Werewolf.

\subsection{Efficiency}

In this subsection, we estimate the number of cards and shuffles
required by our protocol. We break down the cost by protocol type and
by phase of the game.

\paragraph{Role-internal information-sharing protocol.}
For a role-internal information-sharing protocol with $n'$ surviving
players and sharing-information bundles of size $ \hshr $, each player holds
$n'-1$ sharing-information-card bundles (real bundles for sharing
players and dummy bundles for non-sharing players) and one
sharer-identification-card bundle.

%When we construct the grid used in the protocol, we additionally place $n'-1$ dummy sharing-information-card bundles in the lower row for each player. Hence, the total number of cards used for sharing-information bundles (real and dummy) is
%$
%  h n'(n'-1) \quad (\text{upper row})
%  \;+\;
%  h n'(n'-1) \quad (\text{lower row})
%  = 2 h n'(n'-1).
%$

For the information-sharing part, the protocol uses \(n'-1\)
sharing-information-card bundles for each player and \(n'-1\) dummy
bundles in the corresponding dummy row.  Since each bundle has size
\(\hshr\), the information bundles, real and dummy together, use
\(2\hshr n'(n'-1)\) cards.

In addition, the grid contains \(2n'\) sharer-identification-card
bundles, each consisting of two cards, and \(2n'\) player-number cards.
These non-information cards contribute \(4n' + 2n' = 6n'\) cards.

In Step~2, for each player we create a preserved copy of her
sharer-identification commitment.  By the copy primitive, one preserved
copy requires four helper cards in addition to the original commitment.
Thus the copy operations require \(4n'\) temporary helper cards in total.
The subsequent XOR operation combines this copied commitment with the
fixed commitment in the dummy row and requires no additional fresh cards
beyond its input commitments.  Therefore, if these temporary helper cards
are included, the role-internal information-sharing protocol uses
$ 2\hshr n'(n'-1) + 6n' + 4n' = 2\hshr n'(n'-1) + 10n'$
cards in total.

As for shuffles, the protocol performs $n'$ applications of the copy
primitive and $n'$ applications of the XOR primitive, and each of
these applications uses one random bisection cut. In addition, we
perform two pile-scramble shuffles (one in Step~3 and one in
Step~5). Thus, the total number of shuffles is
$ 2n' + 2.$

\paragraph{Action-specification protocol.}
For an action-specification protocol with $n'$ surviving players, each
player is given $n'$ role-processing-card bundles in advance (one
$\heart\club$ bundle and $n'-1$ $\club\heart$ bundles for an
action-specifying player, and $n'$ $\club\heart$ bundles for a
non-action-specifying player), so that the total number of
role-processing cards prepared is
$ 2 n'^2.$
In addition, each player receives one action-processing-card bundle of
size $\hact$, so we need $\hact n'$ action-processing cards in total.

In Step~1 of the protocol, we place one no-target role-processing
bundle, namely a \(\heart\club\) bundle, in each of the first \(n'\)
columns of the first row.  We also place one fixed \(\club\heart\)
bundle in each self-target position \(p_{j+1,j}\) for
\(1 \le j \le n'\), and one dummy action-processing bundle of size
\(\hact\) in the last column of the first row.  This requires
\(4n' + \hact\) extra cards.

For the action-specification protocol, the exact constant depends on the chosen implementation of the multi-input OR operation in Step 2.
Since the OR primitive defined in Section 2.2.3 is a two-input primitive, the OR of the \(n'\) copied commitments in each column can be implemented by a linear number of two-input OR operations.
The copy, OR, and XOR operations in Step~2 are applied only to role-processing commitments, each of which consists of two cards.
Together with the \(n'^2\) copy operations and the \(n'\) XOR operations, this gives \(O(n'^2)\) primitive applications and \(O(n'^2)\) cards, including temporary helper cards.
The action-processing bundles themselves contribute only \(O(\hact n')\) cards.
Therefore, the total number of cards used by the action-specification protocol is \(O(n'^2+\hact n')\).

The optional procedure for verifying the number of players who choose a target adds \(O(n'^2)\) role-processing commitments and \(O(n'^2)\) shuffle operations.
The optional procedure for identifying the original owner of each action-processing bundle adds one player-number column, and hence \(O(n')\) cards, together with only a constant number of additional pile-scramble shuffles.
Therefore, including these optional procedures does not change the asymptotic bound \(O(n'^2+\hact n')\) for the action-specification protocol.

As for shuffles, Step~2 uses $n'^2$ random bisection cuts for the copy primitives, a linear number of random bisection cuts per column for the two-input OR primitives, and $n'$ random bisection cuts for the XOR primitives.
Thus, Step~2 uses $O(n'^2)$ random bisection cuts in total.
Including the pile-scramble shuffles in Steps~3, 4, and 5, the total number of shuffles in the action-specification protocol is $O(n'^2)$.

\paragraph{Attribute-judgment protocol.}
In an attribute-judgment protocol with $n'$ surviving players, each
player is given 
%$n'$ role-value bundles in advance, each bundle consisting of $h$ cards encoding the binary attribute associated withher role, and $n'$ attribute-judgment bundles, each of size~2.
$n'$ role-value-card bundles and $n'$ attribute-judgment-card bundles. 
Each of these bundles consists of two cards in the basic binary implementation described in Section~\ref{sec:Attribute-judgment protocol}.
Hence the pre-distributed cards for this protocol contribute $n'(2n'+2n')(=4n'^2)$ cards in total.
%Thus, each player holds $h n' + 2n'$ cards that may be used in
%attribute judgment and related procedures, 
%%and in total we need $n' (hn' + 2n') = (h+2) n'^2$ such cards.
%and the total number of such cards that we need is $n' (hn' + 2n') = (h+2) n'^2$.

In one execution of the attribute-judgment protocol, all \(n'\) iterations are performed.
Across these iterations, each player uses all \(n'\) role-value-card bundles when that player is the target, and uses one attribute-judgment-card bundle in each iteration.
Thus each player uses \(n'\) role-value-card bundles and \(n'\) attribute-judgment-card bundles in total.
These pre-distributed cards can be reused across different executions and across different nights after they are discarded and re-prepared in a role-independent manner.
The protocol performs one pile-shifting shuffle in each iteration, and hence uses \(n'\) pile-shifting shuffles.
%In a single execution of the attribute-judgment protocol, we use $hn'$ cards from the target player's role-value bundles and one attribute-judgment bundle from each player, but the same pre-distributed cards can be reused across different targets and across different nights. The number of pile-shifting shuffles required by the protocol is $n'$, since we perform one pile-shifting shuffle for each possible target player.

\paragraph{Overall cost for Werewolf.}
Let \(h_{\max}=\max\{\hshr,\hact,2\}\). 
We now combine these estimates to bound the total number of cards used in one game of Werewolf with $n$ players.

\begin{itemize}
  \item In the pre-game phase, the Werewolf introduction protocol uses \(2h_{\mathrm{shr}}n(n-1)+10n\) cards, which is \(O(h_{\mathrm{shr}}n^2)\).

  \item In each night phase, since $n'$ players are still alive, 
    \begin{itemize}
      \item at most \(2h_{\mathrm{shr}}n'(n'-1)+10n'\) cards are used for the attack-target sharing protocol~(i.e., a role-internal information-sharing protocol), 
      \item at most $ O(4 n'^2)$ cards are used for the Seer protocol~(i.e., an attribute-judgment protocol), and
\item at most \(O(n'^2+h_{\mathrm{act}}n')\) cards are used for each of the Bodyguard protection and Werewolf attack protocols~(i.e., an action-specification protocol).
    \end{itemize}
Hence, a single night phase with \(n'\) surviving players uses \(O(h_{\max}n'^2)\) cards in total. 

  \item    
  %暫定
  Since at least one player is eliminated on each day, the
  number of night phases is at most $n-1$. Summing the above bound
  over all night phases and adding the pre-game cost, we obtain that
  the total number of cards that ever appear during one game is
  $O\bigl(h_{\max} n^3\bigr).$
%By reusing discarded cards after appropriate shuffles, it suffices to prepare $O\bigl(\hshr n^3\bigr)$ physical cards in advance.
\end{itemize}
%  By reusing discarded cards after appropriate shuffles, it suffices
%  to prepare $O\bigl((h+1) n^3\bigr)$ physical cards in advance.

  %If we focus only on reducing the number of distinct physical cards, we can go one step further: we may prepare only the cards needed for the first day and, at the beginning of each subsequent day, create fresh copies of these cards using the copy primitive after sufficiently shuffling the discarded cards. In this way, the number of distinct physical cards that need to be prepared in advance can be reduced to $O\bigl(\hshr n^2),$ at the cost of an increased number of shuffles.

  The cumulative number of card appearances over the whole game is \(O(h_{\max}n^3)\).  This does not mean that \(O(h_{\max}n^3)\) distinct physical cards must be prepared in advance.
If cards that are no longer active are recycled in later phases, however, one must ensure that the number or composition of the recycled cards does not leak information about the current role distribution among the surviving players.  For example, before reusing such cards, we may mix them with a constant number of dummy cards drawn from a role-independent distribution unknown to the players, and then apply the appropriate shuffles.  With such padding and recycling, the number of distinct physical cards that need to be prepared in advance can be reduced to \(O(h_{\max}n^2)\), at the cost of an increased number of shuffles.

\subsection{Other roles and extensions}
Finally, we briefly explain how several representative additional roles in
\WW can be implemented using our framework.  The roles discussed
below are not meant to exhaust all known variants; rather, they are chosen
as typical examples illustrating how role-specific abilities can be
incorporated into the same protocol framework.

\paragraph{Witch.}
The Witch has two one-time-use abilities: a poison and a healing
potion.
Immediately after the Werewolves' attack, the Witch learns which
player was attacked (but only the Witch learns this).
The Witch then decides whether to use the poison to kill some player
of her choice, and whether to use the healing potion to save the
attacked player.
We implement the attacker-confirmation protocol as a modification of
the attribute-judgment protocol applied to the AND of the Bodyguard's
and Werewolves' action-processing cards.
The poison protocol and healing protocol are realized as
action-specification protocols in which the Witch is an
action-specifying player only once during the game.

\paragraph{Twins.}
Two players are chosen at the beginning of the game to be Twins.
They know each other's identity, and if one Twin dies, the other
dies immediately as well.
The introductory phase for Twins is implemented as a
role-internal information-sharing protocol, with the Twins as sharing
players.
The follow-up death can be implemented by an AND protocol over the
role-value cards combined with an attribute-judgment-like procedure
that identifies the surviving Twin when one Twin dies.

\paragraph{Cupid and Lovers.}
Cupid chooses two players on the first night and makes them Lovers.
The Lovers win if they are both alive when either the village side or
the Werewolf side would normally win, and in some variants they win
only if all other players die.
Cupid's choice of Lovers can be implemented by two instances of the
action-specification protocol, and the Lovers' mutual knowledge and
follow-up deaths can be implemented in essentially the same way as
for Twins.

\paragraph{Hunter.}
The Hunter may, upon his death, choose one other player to kill.
Since this is a public decision made at the moment of death, it can
be implemented without any card-based protocol: the Hunter simply
announces his target.

\paragraph{Fanatic.}
The Fanatic knows who the Werewolves are, but the Werewolves do not
know the Fanatic.
We can modify the Werewolf introduction protocol so that the sharing
players are the Werewolves and the Fanatic, while only the Werewolves
share their player numbers.
In this way, the Fanatic learns the Werewolves' identities, whereas
the Werewolves cannot distinguish the Fanatic from other roles.

\paragraph{Medium.}
The Medium learns, after each execution, whether the executed player
belonged to the Werewolf side or the village side.
This can be implemented by a variant of the attribute-judgment
protocol applied only to the most recently executed player, with the
Medium as the only attribute-judging player and role values encoding
Werewolf versus non-Werewolf.

\section{Discussion}
The moderatorless \WW protocol can be viewed as an instance of a more general idea: replacing a trusted information manager by public, auditable card operations. 
To indicate the scope of this viewpoint, we briefly discuss two further settings, \GG and second-price auctions.
\subsection{Game of the Generals}
%In this subsection, we briefly discuss other games that typically require a human referee.

Games such as \emph{Game of the Generals} and \emph{Gunjin Shogi} are
board games similar in spirit to \emph{Stratego}, but they cannot be
played according to the standard rules without a referee.
Unlike Stratego, when two pieces meet in these games, the identities
of the pieces are not revealed; only the outcome of the confrontation
is announced to both players, and the losing piece is removed from
the board.

Conceptually, our approach applies to these games as well: it is in
principle possible to eliminate the referee without modifying the
rules.
One can assign an ID number to each piece, as we assign player
numbers in \WW, associate each ID with the corresponding piece type,
and, whenever two pieces collide, copy the types linked to their IDs
and apply Boolean operations to compute and output only the result of
the encounter.

Executing such computations would be substantially more demanding
than in the \WW setting and is not realistic with current
card-based techniques.
However, further progress on efficiency-improving methods for
card-based protocols may eventually make it feasible to carry out
these computations fast enough that players would be willing to use
them in actual play.

\subsection{Second-price auction with card-based zero-knowledge}

In this subsection, we describe a card-based zero-knowledge protocol
for a second-price auction.
A second-price auction is an auction format in which the highest
bidder wins the item but pays the second-highest bid as the price.
Because bidding one's true valuation is often a dominant or at least
a reasonable strategy in this format, it has been widely adopted as
the underlying mechanism for bidding algorithms in search engines and
advertising platforms.

In a second-price auction, each player has a bid as her private input.
The player with the highest bid wins the item and pays the second
highest bid.
Card-based implementations of auctions have already been studied by
Haga et al.~\cite{Mizuki2022SecureSorting}.
Their protocol provides a secure sorting procedure on given inputs:
it sorts pairs of (bid, player number) without turning any card
face-up and thereby realizes the auction.
In their protocol, by revealing the cards at certain positions in the
sorted sequence, the winner's player number and the winning price are
made public to all players.
On the other hand, any information unrelated to the winning price or
the winner's player number remains hidden.

In this paper, we modify Haga et al.'s protocol so that only the
winner learns the winning price and, moreover, players who do not win
do not learn who the winner is.
Our protocol proceeds as follows.
\begin{enumerate}
  \item Using the protocol of Haga et al., we sort the pairs
  (bid, player number).
  We call the cards representing the bids \emph{bid cards} and the
  cards representing the player numbers \emph{player-number cards}.

  \item Among the sorted bundles, we add a \dia\ card to the bundle
  with the largest bid, a \heart\ card to the bundle with the
  second-largest bid, and a \club\ card to every other bundle.
  We refer to these added cards collectively as \emph{mark cards}.

  \item We turn all cards face-down and apply a pile-scramble shuffle.

  \item We turn face-up the mark card in each bundle.
  Using the copy protocol, we copy the bid cards from the bundle
  marked by \heart\ into the bundle marked by \dia.
  Here we use the fact that, in the protocol of Haga et al., bid
  cards are encoded as binary bits using \(\heart\club = 1\) and
  \(\club\heart = 0\), so that they can be copied.
  After the copying step, we replace every revealed \heart\ mark card by a
  \club\ mark card.

  \item We turn all cards face-down again and apply another
  pile-scramble shuffle.

  \item We reveal all player-number cards and hand each bundle to the
  player whose number appears on the player-number card in that
  bundle.
  Each player privately inspects the cards she receives.
\end{enumerate}

In the above protocol, the winner receives the bundle containing the
\dia\ card, and the corresponding bid cards have been replaced by the
second-highest bid.
Therefore, the winner learns that she has won and learns the price
she has to pay.
On the other hand, every non-winning player receives only the cards
she initially contributed together with \club\  cards, and thus
learns nothing beyond the fact that she did not win.

\section{Conclusion}

We have presented a complete moderatorless implementation of Werewolf
using only physical cards.  The protocol replaces the moderator's core
functions by three types of card-based subprotocols: role-internal
information sharing, secret action specification, and attribute judgment.
Together, these subprotocols allow the required hidden information and
night actions to be handled without private communication or digital
devices.

The construction demonstrates that a publicly observable sequence of card
operations can still induce role-dependent views when combined with
players' private role cards.  This suggests a general approach to
removing trusted moderators or referees from hidden-information settings.
We also discussed \GG and second-price auctions as examples, suggesting that the same viewpoint can be useful beyond \WW.  Future work includes further reducing the number of cards and shuffles,
simplifying the protocol for practical play, and extending the framework
to further roles and other settings involving trusted intermediaries.

\bibliographystyle{plainurl}
\bibliography{werewolf}

\end{document}